\documentclass[12pt]{article}
\usepackage[dvips]{epsfig}
\usepackage{latexsym}
\usepackage{caption}  
\newcommand{\be}{\begin{equation}}
\newcommand{\ee}{\end{equation}}

\newcommand{\ind}[1]{\mbox{\tiny{#1}}}

\def\bbox{{\,\lower0.9pt\vbox{\hrule \hbox{\vrule height 0.2 cm
\hskip 0.2 cm
\vrule  height 0.2 cm}\hrule}\,}}

\newcommand{\beq}{\begin{equation}}
\newcommand{\eeq}{\end{equation}}
\newcommand{\bea}{\begin{eqnarray}}
\newcommand{\eea}{\end{eqnarray}}

\begin{document}
\setlength{\unitlength}{1mm}
\title{{\hfill {\small } } \vspace*{2cm} \\
Black Holes with Polyhedral Multi-String 
Configurations}
\author{\\
V. Frolov${}^1$ and  D. Fursaev${}^2$}
\maketitle
\noindent  {
$^{1}${ \em
Theoretical Physics Institute, Department of Physics, \ University of
Alberta, \\ Edmonton, Canada T6G 2J1}
\\ $^{2}${\em Joint Institute for
Nuclear Research,
Bogoliubov Laboratory of Theoretical Physics,\\
 141 980 Dubna, Russia}
\\
e-mails: frolov@phys.ualberta.ca, fursaev@thsun1.jinr.ru
}
\bigskip

\begin{abstract} 
We find exact  solutions of the Einstein equations which describe a
black hole pierced by infinitely thin cosmic strings. The string
segments enter the black hole along the radii and their positions 
coincide with
the symmetry axes of a regular polyhedron. Each 
string produces an angle deficit proportional to its tension, while the
metric outside the strings is locally Schwarzschild one. 
There are
three configurations corresponding to  tetrahedra, octahedra and
icosahedra where the number of string segments is 14, 26 and 62,
respectively. There is also a "double pyramid" configuration where the
number of string segments is not fixed.
There can be two  or 
three independent types of strings
in one configuration. Tensions of strings 
belonging to the same type must be
equal.
Analogous polyhedral multi-string configurations
can be combined with other spherically 
symmetric solutions of the Einstein equations.
\end{abstract}

\bigskip

\baselineskip=.6cm

\newpage

\section{Introduction}
\setcounter{equation}0

Although an exact solution for a static black hole  pierced by a single
straight infinitely thin cosmic string \cite{AFV} is easy to find its
multi string generalizations have not been constructed. Some particular
configurations which could be used to describe a black hole with three
strings were discussed in \cite{DoCh:92}. However, by the symmetry they
required string tensions to be of the order of  unity\footnote{We use
the system of units where $G=c=\hbar=1$. For a string with the physical
tension $\mu$ the dimensionless tension  is defined as $G\mu/c^2$. }
which is too large, perhaps, to  be interesting for  physical
applications.   In this paper we find exact solutions of the  Einstein
equations which represent a black hole pierced by more than one cosmic 
string. The strings enter the black hole along the radii and  their
positions coincide with the symmetry axes of a regular  
polyhedron{\footnote{It is interesting to note that some similar
configurations of defects
related to the symmetries of a polyhedron were discussed
in condensed matter physics \cite{Nicolis}.}.
The
symmetry guarantees an equilibrium of this system. Locally the metric
outside the strings remains the Schwarzschild one.

The motivation for this work is to find non-trivial solutions of black
hole -- multi-string configurations where the backreaction effect 
can be
taken into account explicitly. Although such solutions are interesting
by themselves they can also have an interesting physical application.
Configurations with a large number of strings 
attached to the black hole can be
used for very efficient energy mining from the black hole
\cite{FrFu:2000}. It becomes possible because
the 
Hawking energy emitted in the form of string excitations for each
of the strings is of the same order of magnitude as the bulk Hawking
radiation.

\section{Polyhedral string configurations}
\setcounter{equation}0

Let us first discuss the conditions which are sufficient for an
equilibrium  of a multi-string system. Because the black hole metric is
spherically symmetric   there exists  a special case when one can
guarantee that forces acting on strings vanish identically due to
the symmetry.  It happens when there exists a symmetry transformation,
rotation, which transforms the system (a black hole with attached
strings) into itself.  Indeed, a force acting on the string is always
orthogonal to the string.  Suppose  the system is invariant under a
rotation at some finite angle around the string axes. Then the force
acting on the string must vanish since it should remain invariant under
the rotation. Let us consider such configurations in more detail. To
describe them we first neglect the string tension,  that is consider
strings as test objects.

Since strings are directed along the radii, their positions can be
identified with  points (vertexes) on a unit sphere $S^2$.  Consider
the five Platonic solids. A Platonic solid is a regular polyhedron. Its
faces are regular $p$-gons, $q$ surrounding each vertex. These
configurations are denoted as $\{ p,q\}$. The five Platonic solids are
tetrahedron ($\{ 3,3 \}$), octahedron ($\{ 3,4 \}$), cube ($\{ 4,3
\}$), icosahedron ($\{ 3,5 \}$), and dodecahedron ($\{ 5,3 \}$) (see
e.g. \cite{Coxeter:63}). By projecting the edges of a polyhedron from
its center onto a concentric sphere, we obtain a set of arcs of great
circles on the sphere which intersect at the vertexes. This gives a
spherical tessellation which is also denoted as $\{ p,q\}$. This
spherical tessellation is invariant under rotations around each of its
vertexes at the angle $2k\pi/q$ where $k<q$ are natural
numbers. The system with radial strings which cross a sphere at these
vertexes is invariant under discrete rotations, and hence must be in an
equilibrium. 

The number of strings segments $N_s$ which coincides with the number of
vertexes can be easily found. Denote by $n_v$, $n_e$, and $n_f$ the
number of vertexes, edges and faces for a given spherical tessellation.
They are connected by  Euler's formula 
\be 
n_v-n_e+n_f=2\, . 
\ee 
Since $n_f\, p/2=n_e$ and $n_v\, q/2=n_e$, one has
\begin{equation}\label{n}
n_e= {1\over p^{-1}+q^{-1}-2^{-1}}\, . 
\end{equation} 
\be 
N_s\equiv n_v= {2\over q}\, {1\over p^{-1}+q^{-1}-2^{-1}}\, . 
\ee 
The number $N_s$ of string segments attached to the black hole for the
Platonic configurations are 4 (tetrahedron), 6 (octahedron), 8 (cube),
12 (icosahedron), and 20 (dodecahedron). 

The number of strings in the equilibrium can be made larger. Additional
strings can be attached to the centers of faces and mid-edge points.
Together with original axes of $q$-fold rotations, new vertexes
connected with the center of the sphere are axes of $p$-fold and
$2$-fold rotations, respectively. It can be shown that no further
possible axes of rotation can occur \cite{Coxeter:63}. Thus, the
symmetry operations of the polyhedron consist of rotations through
angles $2 k\pi/q$, $\pi$, and $2j\pi/p$, around these  axes.  It means
that there is a discrete symmetry group which  transforms the string
configuration into itself. Corresponding groups are known to be
symmetry groups of regular polyhedra  (see, for instance,
\cite{Coxeter:63}--\cite{Forsyth}). For $\{ p,q\}$ and $\{ q,p\}$ the
groups are the same. The order of the discrete rotation group is
$2n_e$, and is equal to 12, 24, and 60 for tetrahedral, octahedral and
icosahedral group, respectively.  Let $N_v$, $N_e$, and $N_f$  be the
number of vertexes, edges, and faces of this new spherical
tessellation, then one has
\begin{equation}\label{n2}
N_v= 2n_e+2\, ,\hspace{0.5cm}
N_e= 6n_e\, ,\hspace{0.5cm}
N_f= 4n_e\, .
\end{equation}
Thus, the number of strings segments now is $N_s=N_v$ is 14, 26, and 62 for
tetrahedral, octahedral and icosahedral configuration, respectively. 
Following to \cite{DoCh:92} we call such a configuration {\it
polyhedral strings}. It is important that for such a spherical
tessellation, the faces are isometric spherical triangles, {\it
fundamental triangles}, formed by arcs of great circles on the 
sphere.  Because we have a triangulation of $S^2$ the same results can
be derived by another  method which will be useful when we discuss the
backreaction problem in  the next Section.

Suppose that the angles of the fundamental triangle are 
$\pi\lambda_k$, $k=1,2,3$. 
The area ${\cal A}_\Delta$
of the triangle on the unit sphere is given by
Girard's formula \cite{Coxeter:63}
\begin{equation}\label{1.1}
{\cal A}_\Delta=\pi(\lambda_1+\lambda_2+\lambda_3-1)~~~.
\end{equation}
The triangulation possesses planes of the symmetry and,
hence, the number of fundamental triangles is even, $N_f=2N$.
Because the vertices of the triangles coincide with the symmetry axes
the angles can take only discrete values,
$\lambda_k=1/l_k$, where $l_k$ are positive
integers. 
For the fundamental triangles which cover the 
surface of $S^2$ one time one gets from (\ref{1.1})
the relation
\begin{equation}\label{1.2}
N\left({1 \over l_1}+{1 \over l_2} + {1 \over l_3} -1\right)=2~~~.
\end{equation}
Because $N=2n_e$ one can see that (\ref{n})
follows from (\ref{1.2}) when $l_1=p$, $l_2=q$, $l_3=2$.
The number of strings is $N_s=N+2$, see (\ref{n2}).

Before discussing possible solutions of (\ref{1.2}) let us mention one
more instructive derivation of this equation. Suppose that $N_k$  is
the number of vertexes which join  $2l_k$ edges. There are $2l_k$
fundamental triangles around the vertex of this $l_k$-fold type. They
make a spherical polygon  with the area $2l_k{\cal A}_\Delta$.  Again,
by the symmetry  arguments, the sphere is completely covered by $N_k$
polygons of this type. It follows then that $2l_k N_k{\cal
A}_\Delta=4\pi$ and $N_k=N/l_k$. Thus, by taking into account that
$N_1+N_2+N_3=N_s=N+2$, we get (\ref{1.2}).

It can be shown that there  are only four types of solutions of
(\ref{1.2}) when $l_k$ are natural numbers \footnote{If the sphere is
covered by the fundamental triangles more than once one obtains eleven
more additional configurations of the same four types
\cite{Forsyth}.}.  All these solutions correspond to different
symmetries.   The three solutions are the described Platonic
configurations (a tetrahedron ($l_1=2$, $l_2=l_3=3$, $N=12$), a
octahedron ($l_1=2$, $l_2=3$, $l_3=4$, $N=24$),  and a icosahedron
($l_1=2$, $l_2=3$, $l_3=5$, $N=60$) ).  The fourth solution is a
"double pyramid" where $l_1=l_2=2$ and $l_3=n$  is arbitrary, $N=2n$.
The corresponding string configuration for the double pyramid is
illustrated by  Figure~\ref{f1}.

\begin{figure}
\centerline{\epsfig{file=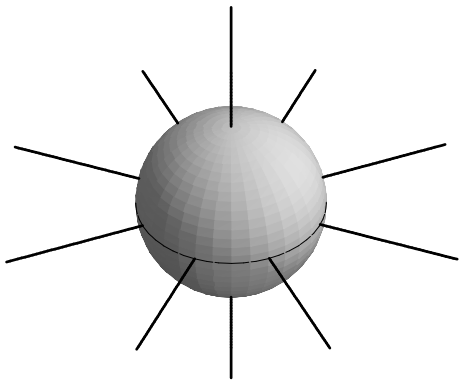, width=6cm}}
\caption[f1]
{A double pyramid configuration with $N_s=10$}
\label{f1}
\end{figure}

\section{Backreaction}
\setcounter{equation}0

Consider now the gravitational backreaction effect for polyhedral
strings.  For a finite string tension there exists an angular deficit
around the string which should be taken into account.  Also, if
strings have different and arbitrary tensions the symmetry
arguments used above may not work.

Let us begin with the simplest and physically most interesting 
case when strings have equal tensions which result in a  deficit angle
$\alpha$. We assume that each of the angles of the fundamental
triangle  changes by the same factor $1-\alpha$. Girard's formula gives
for the area of the fundamental triangle
\begin{equation}\label{1.4}
{\cal A}_\Delta=\pi\left[(1-\alpha)
\left({1 \over l_1}+{1 \over l_2}+{1 \over l_3}\right)-1\right]~~~.
\end{equation}
To get our configuration we take a number of these triangles and glue
them together along the edges in the same way as one glues the fundamental
triangles on the ordinary $S^2$. Note that the edges of the triangles
are still segments of great circles. This guarantees that the extrinsic
curvature on the edges  is zero and any two triangles are joined 
without jumps in the extrinsic curvature.    

Suppose that by gluing $2N$ fundamental triangles together  we obtain
the triangulation of $S^2$. Such a surface $S^2_{\ind{cone}}[N_s]$ 
has  $N_s=N+2$ cone-like singular points with the angle deficits
$\alpha$.  We demonstrate now that sufficient condition for
existence of such a triangulation of  $S^2_{\ind{cone}}[N_s]$ again
takes the form (\ref{1.2}).  First, let us find the total area $\cal A$
of $S^2_{\ind{cone}}[N_s]$. This can be done by using the Gauss-Bonet
formula. For $S^2_{\ind{cone}}[N_s]$ with deficit angles $\alpha_s$
this formula takes the form (see, e.g., \cite{FS})
\begin{equation}\label{1.3a}
2\cdot 4\pi=\int d^2x \sqrt{\gamma} R+4\pi\sum_s\alpha_s=
2{\cal A}+4\pi\sum_s \alpha_s~~,
\end{equation}
where $R=2$ is the scalar curvature and the integration goes over the
smooth domain where the metric coincides with the metric on $S^2$. From
(\ref{1.3a}) one finds
\begin{equation}\label{1.3}
{\cal A}
=2\pi\left(2-\sum_s \alpha_s\right)~~~.
\end{equation}
In the considered case ${\cal A}=\pi(2-N_s\alpha)$ where $N_s$ is
the total number of strings. Now, if $2N$ triangles cover the surface
completely one gets the relation
\begin{equation}\label{1.5}
N\pi\left[(1-\alpha)
\left({1 \over l_1}+{1 \over l_2} + {1 \over l_3}\right) -1\right]=
\pi(2-\alpha(N+2))~~~.
\end{equation}
After a simple transformation (when the term $-\pi \alpha N$
is taken from the left hand side to the right)
it reduces to (\ref{1.2}) and proves that our triangulation is  
self-consistent.

We can generalize this result to the case when  different types of the
symmetry axes  correspond to strings with different tensions. This will
be consistent with the symmetry group. Thus, in general for
configuration with three different axes one can have three types of
strings with tensions $\mu_k$ ($k=1,2,3$) including the case when some
$\mu_k$ can  vanish\footnote{To avoid the confusion let us emphasize
that if the two angles of the fundamental triangle coincide they
correspond to strings with the same tension. This is needed to preserve
the symmetry of the problem.}.   Now the triangles have
the angles $\pi(1-\alpha_k)/k$ where $\alpha_k=4\mu_k$   are the
corresponding angular deficits around the strings. As in the previous
case, we take $2N$ such triangles and glue them along the edges to get
triangulation equivalent to triangulation of $S^2$ by the fundamental
triangles.  Let us show that this triangulation is selfconsistent and
agrees with (\ref{1.2}) for any $\mu_k$.  As  follows from (\ref{1.3}),
the area of the sphere is
\begin{equation}\label{1.6}
{\cal A}=2\pi\left(2-N_1\alpha_1-N_2\alpha_2-N_3\alpha_3\right)~~~.
\end{equation}
The area of the fundamental triangle is 
\begin{equation}\label{1.7}
{\cal A}_\Delta=\pi\left[(1-\alpha_1){1 \over l_1}+
(1-\alpha_2){1 \over l_2}+(1-\alpha_3){1 \over l_3}
-1\right]~~~.
\end{equation}
Therefore, because $2N{\cal A}_\Delta={\cal A}$,
\begin{equation}\label{1.8}
N\left[(1-\alpha_1){1 \over l_1}+
(1-\alpha_2){1 \over l_2}+(1-\alpha_3){1 \over l_3}
-1\right]=
\left(2-N_1\alpha_1-N_2\alpha_2-N_3\alpha_3\right)~~~.
\end{equation}
If we now take into account relation between the number of vertexes and
the number of triangles, $N/l_k=N_k$, (\ref{1.8}) will be reduced to 
(\ref{1.2}). Thus, the values $\alpha_k$ can be chosen  
arbitrary. In particular, some of them can be zero, which means that
there are no strings along the corresponding symmetry axes.
When only one parameter $\mu_k\neq0$ for some $l_k\neq 2$
one gets a configuration of strings which go through the vertexes
of a Platonic solid $\{p,q\}$ with $p=l_k$. The double pyramid
in this case is reduced to a single straight string.

\section{Derivation of solutions}
\setcounter{equation}0

Let us discuss now solutions of the Einstein equations for the
polyhedral string configurations. The total action of the system 
is\footnote{In this section we restore a
normal value $G$ for the Newton constant.}
\begin{equation}\label{1.9}
I={1 \over 16\pi G}\left[\int_{\cal M} \sqrt{-g}d^4x R+
2\int_{\partial {\cal M}}K\sqrt{-h}d^3x\right]-
{1 \over 4\pi}\sum_s \mu_s\int\sqrt{\sigma}_sd^2\zeta_s~~~.
\end{equation}
The last term in the right hand side of (\ref{1.9})  is the Nambu-Goto
action of string, where $(\sigma_s)_{\alpha\beta}$ 
is the metric induced on the
world-sheet of a particular string. We assume in (\ref{1.9}) that the
space-time $\cal M$ has a  time-like boundary.  We take the metric in
the form
\begin{equation}\label{1.10}
ds^2=\gamma_{\alpha \beta}dx^\alpha dx^\beta+e^{2\phi}
a^2d\Omega^2~~~.
\end{equation}
Here $\gamma_{\alpha\beta}$ is a 2D metric, $\phi=\phi(x)$ a dilaton
field which depends on coordinates $x^\alpha$, and $d\Omega^2$ is the
metric on $S^2_{\ind{cone}}[N_s]$. For a string located at fixed angles
the induced metric on  a string world-sheet coincides with 
$\gamma_{\alpha\beta}$. The parameter $a$ in (\ref{1.10}) has the
dimensionality  of the length. Locally near each string the metric
$d\Omega^2$ can be written as
\begin{equation}\label{1.11}
d\Omega^2=\sin^2\theta d\varphi^2+d\theta^2~~~,
\end{equation}
where $0\leq \theta \leq \pi$, and $\varphi$ is periodic
with period $2\pi(1-\alpha_s)$.
To proceed we have to take into account in (\ref{1.9}) 
the presence of 
delta-function-like contributions due to the conical 
singularities \cite{FS}
\begin{equation}\label{1.13}
\int_{\cal M} \sqrt{-g}d^4x R=
\int_{\cal M'} \sqrt{-g}d^4x R+4\pi\sum_s\alpha_s \int 
\sqrt{\sigma}_sd^2\zeta_s~~,
\end{equation}
where $\cal M'$ is the regular domain of $\cal M$.
If we impose the on-shell condition $\alpha_s=4\mu_s G$
the contribution of the conical singularities in the
curvature in (\ref{1.9}) will cancel exactly the 
contribution from the string actions.
There will remain only the bulk
part of the action.
On the metric (\ref{1.10}) it will reduce to
2D dilaton gravity 
\begin{equation}\label{1.14}
I={1 \over 4G_2} \left[\int \sqrt{\gamma}d^2x\left(
e^{2\phi}R_2+2e^{2\phi}(\nabla\phi)^2+{2 \over a^2}\right)
+2\int dy e^{2\varphi}(k-k_0)\right]~~~,
\end{equation}
\begin{equation}\label{1.15}
{1 \over G_2}={a^2 \over G}C(\mu_s)
\end{equation}
\begin{equation}\label{1.15b}
C(\mu_s)=1-2G(N_1\mu_1+N_2\mu_2+N_3\mu_3)~~~.
\end{equation}
The curvature $R_2$ in (\ref{1.14}) is the 2D curvature
determined by $\gamma_{\alpha\beta}$.
As a result of modification of the area of sphere due to
conical singularities, see (\ref{1.6}),
the gravitational action (including the boundary term)
acquires the overall coefficient 
which depends on $\alpha_s$.
We included this coefficient in the
definition of effective two dimensional gravitational 
coupling $G_2$, Eq. (\ref{1.15}).
It is important that action (\ref{1.14}) has 
precisely the same form as the
dilatonic action obtained under spherical reduction of the
gravitational action in the absence of cosmic strings.
Therefore strings have no effect on the dynamical
equations for the metric $\gamma_{\alpha\beta}$ and the
dilaton $\phi$. For these quantities one has 
standard solutions. In particular, the Birkhoff theorem can be 
applied in this case and guarantees that in the absence of 
the other matter in the bulk, the solution is static
and it is a 2D black hole
\begin{equation}\label{bh}
d\gamma^2=-Fdt^2+F^{-1}dr^2~~,~~~F=1-{2M \over r}~~~
\end{equation}
of the mass $M$. The corresponding four-dimensional solution is a
Schwarzschild black hole of the same mass. In a similar way, by using
(\ref{1.14}) one can  construct non-static solutions in the presence of
polyhedral strings. Non-vacuum static spherically symmetric solutions,
such as a charged black hole with strings, can be constructed as well
by adding a matter in the bulk.

\section{Discussion}
\setcounter{equation}0

The aim of this work was to find exact static solutions which describe
an equilibrium configuration of a black hole    with radial infinitely
thin cosmic strings attached to it. We found special class of such
solutions where the positions of the radial strings  are fixed by the
symmetry.

Are the obtained polyhedral multi-string solutions unique static solutions
for black-hole--multi-string configurations? Certainly not. 
Let us take two equal  spherical triangles and
glue  them along the coinciding edges. We get a configuration  of three
cosmic strings of certain tensions which is not related to any
symmetry.  If the triangles are fundamental one gets a configuration
discussed in \cite{DoCh:92}. It is not clear, however, whether one  can
generalize this procedure and obtain a closed 2D surface with conical
singularities by gluing a certain set of different spherical triangles.
To find a general static multi-string configuration
attached to a black hole is an interesting open
problem.  

Another interesting problem is to investigate 
the possibility of equilibrium configurations of
higher-dimensional defects (like domain walls) and the 
effect of the gravitational backreaction in these systems.
Some configurations of domain walls corresponding to spherical
tessellations are discussed in \cite{BB}.

\noindent
\section*{Acknowledgments}

\indent This work was partially supported  by the Natural Sciences and
Engineering Research Council of Canada and by the NATO Collaborative
Linkage Grant CLG.976417. D.F. was also supported by RFBR grant
N99-02-18146. V.F. is grateful to the Tokyo Institute of Technology for
the hospitality during the preparation of this paper for publication.


\newpage

\begin{thebibliography}{9}


\bibitem{AFV} M. Aryal, L. Ford and A. Vilenkin,
Phys. Rev. {\bf D34} (1986) 2263.

\bibitem{DoCh:92} J.S. Dowker and P. Chang, Phys. Rev. {\bf D46}
(1992) 3458.

\bibitem{Nicolis} S. Nicolis, R. Mosseri, J.F. Sadoc,
Europhys. Lett. {\bf 1} (1986) 571.

\bibitem{FrFu:2000} V.P. Frolov and D.V. Fursaev,
{\it Mining Energy from a Black Hole by Strings}, hep-th/0012260.

\bibitem{Coxeter:63} H.S.M. Coxeter, {\it Regular Polytopes},
The Macmillan Company, New York, 1963.


\bibitem{Coxeter:74} H.S.M. Coxeter, {\it Regular Complex Polytopes},
Cambridge University Press, 1974.

\bibitem{Forsyth} A.R. Forsyth, {Theory of Functions of a
Complex Variable}, Cambridge University Press, 1900.


\bibitem{FS} D.V. Fursaev and S.N. Solodukhin,
Phys. Rev. {\bf D52} (1995) 2133.


\bibitem{BB} D. Bazeia and F.A. Brito, Phys. Rev. {\bf D62}
(2000) 101701, hep-th/0005045.


\end{thebibliography}
\end{document}